\documentclass[openany]{nature}

\usepackage{epsfig}
\usepackage{color}
\usepackage{graphicx}
\usepackage{longtable}
\usepackage{amssymb}
\usepackage{rotating}
\usepackage{gensymb}
\usepackage{textcomp}
\usepackage{amsmath,bm}
\usepackage{ulem}
\usepackage{lineno}

\newcommand{\apj}{Astrophys. J.}

\newcommand{\pasp}{Publ. Astron. Soc. Pac.}
\newcommand{\apjs}{Astrophys. J. Suppl. Ser.}

\newcommand{\mnras}{Mon. Not. R. Astron. Soc.}
\newcommand{\apjl}{Astrophys. J. Let.}
\newcommand{\aap}{Astron. Astrophys.}
\newcommand{\aj}{Astron. J.}
\newcommand{\nat}{Nature}

\newcommand{\natco}{Nature Communications}

\newcommand{\aar}{Astronomy and Astrophysics Review} %{A\&ARv}

\newcommand{\oiii}{\hbox{[O\,{\sc iii}]}}
\newcommand{\nii}{\hbox{[N\,{\sc ii}]}}
\newcommand{\sii}{\hbox{[S\,{\sc ii}]}}

\title{Multiple gas acquisition events in galaxies with dual misaligned gas disks}
\author{Xiao Cao$^{1,2,3}$,
Yan-Mei Chen$^{1,2,3}$,
Yong Shi$^{1,2,3}$,
Min Bao$^{1,2,3,4}$, 
Alexei Moiseev$^{5,6}$, 
Dmitry Bizyaev$^{7,6}$, 
Song-Lin Li$^{1,2,3}$, 
Jos\'e G. Fern\'andez-Trincado$^{8}$, 
Rogemar A. Riffel$^{9,10}$, 
Rogério Riffel$^{11,10}$, 
Richard R. Lane$^{12}$.
}

\begin{document}
\maketitle

\begin{affiliations}
    \item School of Astronomy and Space Science, Nanjing University, Nanjing, China.
    \item Key Laboratory of Modern Astronomy and Astrophysics (Nanjing University), Ministry of Education, Nanjing, China. 
    \item Collaborative Innovation Center of Modern Astronomy and Space Exploration, Nanjing, China. 
    \item School of Physics and Technology, Nanjing Normal University, Nanjing, China. 
    \item Special Astrophysical Observatory, Russian Academy of Sciences, Nizhny Arkhyz, Russia. 
    \item Sternberg Astronomical Institute, Lomonosov Moscow State University, Moscow, Russia. 
    \item Apache Point Observatory and New Mexico State University, Sunspot, NM, USA.
    \item Instituto de Astronom\'ia, Universidad Cat\'olica del Norte, Antofagasta, Chile.  
    \item Departamento de F\'isica, CCNE, Universidade Federal de Santa Maria, Santa Maria, Brazil. 
    \item Laborat\'orio Interinstitucional de e-Astronomia - LIneA, Rio de Janeiro, Brazil. 
    \item Departamento de Astronomia, Instituto de Física, Universidade Federal do Rio Grande do Sul, Porto Alegre, Brazil. 
    \item Centro de Investigaci\'on en Astronom\'ia, Universidad Bernardo O'Higgins, Santiago, Chile. 
\end{affiliations}

\begin{abstract}
Frequent accretion of external cold gas is thought to play an important role in galaxy assembly. 
However, almost all known kinematically misaligned galaxies identify only one gas disk that is misaligned with the stellar disk, 
implying a single gas acquisition event. 
Here we report a new configuration in two galaxies 
where both contain two gas disks misaligned with each other and also with the stellar disk. 
Such systems are not expected to be stable or long-lasting, 
challenging the traditional picture of gas accretion of galaxies and their angular momentum build-up. 
The differences in kinematic position angles are larger than \bm{$120^\circ$} between the two gas disks, 
and \bm{$40^\circ$} between each gas disk and the stellar component. 
The star formation activity is enhanced at the interface of the two gas disks 
compared with the other regions within the same galaxy. 
Such systems illustrate that low-redshift galaxies can still experience multiple gas acquisition events, 
and provide a new view into the origins of galactic gas.
\end{abstract}

\clearpage

\noindent{\large\bf Introduction}

In the framework of hierarchical structure formation theory, a galaxy grows from primordial density fluctuations and its subsequent evolution is regulated by 
a series of internal (for example, stellar winds, supernova exploration, active galactic nucleus (AGN) feedback) and external (for example, gas accretion, merger and interaction) processes. Kinematically 
decoupled galaxies, which host two components (gas or stars) rotating in very different directions with respect to each other, are believed to be key 
laboratories to investigate the influence of external processes on galaxy evolution\cite{Rubin94, Schweizer98}.

Kinematical misalignments in the context of galaxies can occur in a variety of forms, including: (1) gas-versus-stars, where the gaseous disk and the stellar body of a galaxy rotate in 
different directions; (2) stars-versus-stars, where a galaxy hosts two stellar components with different rotation directions; (3) gas-versus-gas, where a galaxy 
hosts two misaligned gaseous disks. The phenomenon of gas and star misalignment is now found to be ubiquitous in elliptical and lenticular galaxies\cite{Bettoni84,Bertola88,Bertola1988,Kuijken96,Kannappan01,Pizzella04,Davis11}, 
and samples of star-forming gas-star misaligned galaxies are also reported thanks to the development of large integral-field spectroscopy surveys\cite{Chen16,Jin16,Martinsson18,Bryant19}. 
Cases of star-star misalignment are not as common as gas-star misalignment; 
statistic studies give an incident rate of $\sim$2\% (refs.\cite{Bevacqua22, Min22}) based on samples from the Mapping Nearby Galaxies at Apache Point Observatory (MaNGA) Survey\cite{Bundy15,Drory15,Wake17}. 
So far, the incidence of gas-gas misaligned galaxies is largely unknown. 
Less than ten gas-gas misaligned galaxies have been reported, 
and observations have been based primarily on long-slit spectroscopy from about two decades ago\cite{Rubin94,Braun94,Fisher94,Plana96,Hunter99,Corsini02}. 
The gas-gas misaligned galaxies either lack observations of stellar kinematics because of poor continuum detections, 
or have one gas disk corotating with the stellar component while the other one is misaligned. 
Similar to gas-star and star-star misaligned cases, the gas-gas misaligned galaxies in the literature can only indicate a single gas acquisition event. 
The low incidence of gas-gas misalignments is due to the fact that the collision cross-section between the two 
gas components is large, leading to a short dynamical friction timescale of about several hundred million years\cite{Roy95,deAvillez02}.

In this work, we investigate two gas-gas misaligned galaxies observed with fibre-optic integral-field units (IFU) in the MaNGA survey, 
finding that in each galaxy the two gas disks 
are misaligned with each other, and that both of them are also misaligned with the stellar component. 
The star formation is enhanced at the interface of the two gas disks, indicating that these galaxies have experienced multiple gas acquisition events;
the interaction between the two gas disks leads to the formation of new stars. 

\noindent{\large\bf Results} 

\noindent{\bf Selection of the gas-gas misaligned galaxies}\\
We analyse gas and stellar velocity fields of a representative sample of 9,456 unique galaxies in internal MaNGA Product Launch-10 (MPL-10). 
Figure 1 shows the two gas-gas misaligned galaxies studied in this work. 
Fig. 1a-i shows the example of one galaxy. 
Figure 1a shows the Sloan Digital Sky Survey (SDSS) false-colour image in which the purple hexagon marks the region covered by the MaNGA bundle. 
The stellar and gas velocity fields are presented in Fig. 1b,c, where the gas velocity field is traced by H$\alpha$ emission. 
Figure 1d shows the kinematic position angle of gas ({\it PA}$_{\rm gas}$, red) and stellar ({\it PA}$_{\rm \star}$, black) components as well as their $\pm1\sigma$ errors as a function of radius. 
We calculate the standard deviation of kinematic {\it PA}s of the gas ($\sigma_{\rm PA,gas}$)  
and stellar ($\sigma_{\rm PA,\star}$) components. 
A parent sample of $\sim$84 objects are selected as $\sigma_{\rm PA,gas}>20^\circ$ and $\sigma_{\rm PA,\star}<5^\circ$.  
These empirical cuts on $\sigma_{\rm PA,gas}$ and $\sigma_{\rm PA,\star}$
include $\sim$85\% of galaxies with disturbed gas kinematics identified through visual classification, 
while at the same time removing contaminations, such as on-going mergers, galaxies with irregular morphologies and groups within a single MaNGA bundle. 
The two galaxies studied in this work are notable in the parent sample in that 
(1) they host two galactic-scale misaligned gas disks and  
(2) both gas disks are misaligned with the stellar disk. 
The evidence for two gas disks can be found as a dramatic change of {{\it PA}$_{\rm gas}$} along the radius in Fig. 1d, 
where the kinematic {\it PA} of the inner (outer) gas disk is marked as a black dotted (solid) line in the gas velocity field. 
Figure 1g shows the gas velocities along the major axes of the inner (orange) and outer disks (cyan).
To quantify the two gas disks, the circular disk rotation field\cite{Krajnovi06} is applied to describe their line-of-sight gas velocity fields, 
as shown in Fig. 1e,h. 
The final kinematic model of two misaligned gas disks is shown in Fig. 1f. 
The figure shows a combination of the inner and outer disk models with the flux ratio as a free parameter.
Figure 1i shows the residual of the gas velocity field. 
The second galaxy is shown in Fig. 1j-r.

We match the MaNGA sample with the literature catalogue 
based on Galaxy Evolution Explorer (GALEX), SDSS and Wide-field Infrared Survey Explorer (WISE) observations 
(GALEX-SDSS-WISE Legacy Catalog (GSWLC)\cite{Salim16}) to obtain the global stellar mass ($M_\star$) and star formation rate (SFR) for 8,012 out of 9,456 galaxies. 
Figure 2 shows the SFR versus $M_\star$ diagram, from which the background contours are for the GSWLC sample, 
the grey dots represent the MaNGA sample and the open triangles mark the 84 galaxies with disturbed gas kinematics. 
The two gas-gas misaligned galaxies are shown as yellow and blue stars. 
There are two clear density peaks in the contours. The top peak with high SFR corresponds to the star-forming 
main sequence, while the bottom peak with little star formation is related with the quiescent sequence (QS), and the green-valley (GV) galaxies are located between these two extremes. 
These three galaxy populations are separated by the two lines on the figure. 
The cyan solid line is an approximation of the lower boundary of star formation\cite{Chang15} (at the $\sim$1$\sigma$ level in scatter), 
while the orange dashed line is an approximation of the upper boundary of the QS\cite{Salim16}. 
The two gas-gas misaligned galaxies are located in the GV.

\noindent{\bf Enhanced star formation at the interface of two misaligned gas disks}\\
In this section, we apply the traditional $\nii$ Baldwin-Phillips-Terlevich (BPT) diagnostic diagram\cite{Baldwin81} to classify the ionization sources for the galaxies. 
Figure 3a-g shows one galaxy as an example. 
Figure 3a shows the $g$,$r$,$z$-band composite image ($\sim$2 mag deeper than SDSS) from the Dark Energy Spectroscopic Instrument (DESI) Legacy Imaging Surveys\cite{Dey19}. 
Figure 3g shows the \oiii$/$H$\beta$ $-$ \nii$/$H$\alpha$ diagram for spaxels with signal-to-noise (S/N) ratio larger than 3 for all the four emission lines. 
For the outer gas disk regions with lower S/N ratio of the emission lines, we stack the spectra to improve the S/N ratio. The two stacked regions of each galaxy are marked in the spatially resolved BPT diagrams (Fig. 3f)
with black and magenta lines, and the line ratios measured from the stacked spectra are shown in Fig. 3g as black and magenta diamonds.
For both the BPT and spatially resolved BPT diagrams, red represents the Seyfert region, blue is the composite of the AGN and star formation, 
orange represents the Low-Ionization Emission-line Region (LIER) according to the demarcation lines given in refs.\cite{Kewley01,Kauffmann03}. 
The whole galaxy is dominated by LIER regions. 
The two pieces of composite area marked by black slashes in the gas velocity field (Fig. 3b) are located at the interface of the two gas disks, which is defined as the transition region 
of {\it PA}$_{\rm gas}$ around the major axes of the two gas disks (the region between $r_1$ and $r_2$). It is also consistent with the area 
between the radius with the maximum gas velocity of the inner disk and where the gas velocity of the outer disk becomes flat (Fig. 1g). 
The interface areas (marked by black slashes in Fig. 3d,e) show a higher H$\alpha$ equivalent width (EQW$_{\rm H\alpha}$) and a lower 4,000{\AA} break (D$n$4000) than the other regions 
within the same galaxy. Considering that the D$n$4000 is a good indicator of the light-weighted stellar population age, lower D$n$4000 at the interface suggests younger stellar populations.
Although H$\alpha$ emission can be contributed by both star formation and black hole activities, there is no doubt that composite regions have a higher fraction of star formation contribution 
to the nebular emission lines relative to the Seyfert and LIER regions, and thus the higher H$\alpha$ EQW at the interface implies stronger star formation activities. 
Figure 3c shows the gas velocity dispersion map, in which again the black slashes mark the regions with enhanced H$\alpha$ EQW.  
It is clear that the gas velocity dispersions at the shaded positions are lower than in the outer regions.

\noindent{\bf Gas-phase metallicity}\\
To have an idea about the origin of the external gas, we analyse the DESI images of these two galaxies following the literature method\cite{LiSL21}, 
finding that there are no remnant features from merging and/or interacting process, including faint tidal streams and/or shells, extended asymmetric haloes and distortion induced by companions.
This result suggests that the acquisition of cold gas from the circumgalactic medium and/or intergalactic medium or gas-rich satellites is the origin of the misaligned gas\cite{Sanchez14,WangS22}. 
One possible way to find out whether the external gas comes from dwarf satellites, the pristine gas in the intergalactic medium or recycled metal-enriched gas in the circumgalactic medium is to 
measure the gas-phase metallicity\cite{Hwang19,Sanchez14b}. We use \nii$\lambda$6585$/$\sii$\lambda\lambda$6718,32 as the indicator of gas-phase metallicity\cite{Kewley02,Smirnova22}. 
Figure 4a,c shows the distribution of gas-phase metallicity for spaxels with the S/N ratio of \nii$\lambda$6585 and \sii$\lambda\lambda$6718,32 larger than 3 for the two galaxies. 
In Fig. 4a, there is evidence that the outer gas disk (marked by blue ellipses) has a gas-phase metallicity of $\sim$8.8, which is $\sim$0.2dex lower than that for the inner disk and $\sim$0.1dex lower than in the 
interface regions. In Fig. 4c, we failed to get the gas-phase metallicity for the outer disk because of the low S/N ratio of the emission lines. 
To get a robust measurement of the gas-phase metallicity for the outer gas disks, we stack the spectra of the outer disks to improve the S/N ratio of the emission lines.
In Fig. 4b,d, we add the gas-phase metallicity measured from stacked spectra. Again, there is a trend that the gas-phase metallicity increases from the outer to the inner disk, although the difference is not as large 
as that shown in Fig. 4a. We note that the typical metallicity gradient of GV galaxies with similar stellar mass is $\sim$0.06dex per effective radius ({\it R}$_{\rm e}$). 

The difference in gas-phase metallicity between the inner and outer disks is not high enough for us to distinguish where the external gas comes from.
The lack of obvious difference in metallicity between the interface of two gas disks and the inner disk can be due to two reasons: 
(1) mixing between the accreted and metal-enriched pre-existing gas; (2) enrichment from star formation. 
In a ``closed-box" model\cite{Dalcanton07}, the metallicity mainly depends on the gas mass fraction $f_{\rm gas} \equiv M_{\rm gas}/(M_{\rm gas}+M_{\rm \star})$, 
where $M_{\rm gas}$ represents gas mass, 
so the abundances get elevated instantaneously as the available gas turns into stars: the low D$n$4000 and high H$\alpha$ EQW at the interface indicate that such stars exist.
This result is consistent with a simulation work based on IllustrisTNG\cite{Khoperskov21}, in which the origin of 25 star–star counter-rotators is studied. 
The results there indicate that highly misaligned accretion of external gas leads to counter-rotating phenomena. The infalling gas is mixing with the pre-existing 
gas to form in situ a counter-rotating gas disk and a counter-rotating stellar disk. 
Owing to the different degree of gas mixing and the 
enrichment from star formation, the counter-rotating stellar component within the central 2kpc has a higher metallicity than that in the outer region. 

\noindent{\large\bf Discussion}

The observational results in this paper provide compelling evidence for multiple accretion events in these two galaxies, forming two misaligned gas disks both of which are also 
misaligned with the stellar component. 
The star formation activity is enhanced at the interface of the two gas disks. 
Figure 5a shows a three-dimensional (3D) cartoon describing the configuration of the two gas disks, 
in which red and blue represent the redshift and blueshift parts of gas components. The interface regions are marked in cyan. 
Figure 5b shows the flux ratio between the inner disk ($F_{\rm in}$) and the 
inner plus outer disk ($F_{\rm tot}\equiv F_{\rm in}+F_{\rm out}$) along the major axes of both the inner (orange) and outer (green) disks from our kinematic model fitting (see {\it Methods} and Fig. 6). 
It is clear that the inner disk is dominant within $r_1$ 
(contributing more than 80\% flux) while the outer disk is dominant outside $r_2$.

Although the warped disk and line-of-sight spiral arms could result in changes of {\it PA}$_{\rm gas}$  along the radius\cite{Battaglia06, Jozsa07, Schmidt14, Shetty07}, 
we prefer to explain the gas kinematics of these two galaxies as dual misaligned disks for the following reasons. 
First, the gas component not only changes its value but also the direction along the semi-major axis also changes.
In Fig. 1g, at $r>0$, the gas velocity varies from around 40 km s$^{-1}$ to around $-$80 km s$^{-1}$, 
indicating the direction of angular momentum of the gas disk changes from the inner to the outer region, 
which is hard to explain with gas warps or spiral arms. 
Second, all the cases of galaxies with warped disks in the literature do 
not show enhanced star formation at the warped area, 
which is similar to the definition of the interface between the two gas disks in our galaxies.

It remains a puzzle why the interface region has enhanced star formation but with lower gas velocity dispersion.
Although we do not fully understand the mechanisms that lead to low gas velocity dispersion at the interface, 
we have to point out that this is a common phenomenon in the MaNGA star-forming galaxy sample: regions with higher H$\alpha$ EQW have lower gas velocity dispersion. 
It would be interesting and worthwhile exploring the mechanisms in a future work with a large MaNGA sample.

In summary, we find two galaxies with dual misaligned galactic-scale gas disks. 
Both gas disks are also misaligned with the stellar component, 
indicating that the galaxies have undergone multiple gas acquisitions. 
The star formation activity at the interface of the two gas disks is enhanced in comparison with the other regions within the same galaxy. 
Such systems, on the one hand, provide direct evidence of multiple gas acquisition events in low-redshift galaxies 
and show a new insight into the origins of galactic gas and, on the other hand, supply a new trigger mechanism for star formation activity.

\begin{methods}

\noindent{\bf Observations and data reduction}\\
The data used in this work comes from the MaNGA survey\cite{Bundy15,Drory15}, 
which provides spatially resolved spectroscopy for an unprecedented sample of $\sim$10,000 nearby galaxies with a flat mass distribution in the range 9 $<$ log $M_\star/M_\odot$ $<$ 11.5 and a median redshift of $\langle z\rangle \approx$ 0.03 (ref.\cite{Wake17}). 
Using the Baryon Oscillation Spectroscopy Survey spectrographs\cite{Smee13} on the 2.5-{\it m} Sloan Telescope\cite{Gunn06,Blanton17}, 
MaNGA uses dithered observations with 17 fiber-bundle integral-field units that vary 
in diameter from 12$''$ (19 fibres) to 32$''$ (127 fibres). 
The spectra have a wide wavelength coverage of 3,600$-$10,300\ \AA\ at a resolution of {\it R} $\approx$ 2,000. 
The $r$-band S/N ratio is 4$-$8 (\AA$^{-1}$ per 2$''$ fiber) at the outskirts of galaxies. 
    
The observables used in this work, including gas and stellar velocities ({\it V}$\rm _{H\alpha}$ and {\it V}$_\star$), 
gas velocity dispersion ($\sigma \rm _{H\alpha}$), H$\alpha$ EQW, D\bm{$n$}4000 and emission-line flux, 
are taken from the MaNGA data analysis pipeline\cite{Westfall19}. 
This pipeline applies the penalized pixel-fitting (pPXF) method \cite{Cappellari04} and a subset of stellar templates from the MaStar library\cite{Yan19} to fit the stellar continuum in each spaxel, 
producing estimates of the stellar kinematics. 
Ionized gas kinematics, as well as the flux, are estimated by fitting a single Gaussian to each emission line after stellar continuum subtraction. 
In this process, all the emission lines are tied together to the same velocity. 
We correct the observed velocity dispersion ($\sigma_{\rm obs}$) to remove the effect of instrumental resolution ($\sigma_{\rm inst}$) as $\sigma_{\rm H\alpha} = \sqrt{\sigma_{\rm obs}^2 - \sigma_{\rm inst}^2}$.  

\noindent{\bf Global SFR and $\bm{{\it M}_{\bf \star}}$}\\
Combining the SDSS and GALEX survey, 
the ultraviolet-optical spectral energy distributions (SEDs) covering a wavelength
$\lambda$ = 0.15$–$0.9$\mu$m 
have been created for a sample of $\sim$700,000 galaxies\cite{Salim16}. 
Using the Code Investigating Galaxy Emission (CIGALE)\cite{Noll09}, they then generate the dust attenuation modified stellar SEDs 
and perform Bayesian SED fitting, 
which producing SFR and $M_\star$, as well as their errors, by building the probability distribution function of the physical parameters. 

\noindent{\bf Measurements of kinematic position angle}\\
To get the kinematic {\it PA} as a function of radius, 
we fit the spaxels within a series of ellipses along photometric major axis with a 0.5$''$ step. 
The kinematic {\it PA} within a certain radius is measured based on the established methods\cite{Krajnovi06}; 
it is defined as the counter-clockwise angle between north and a line that bisects the velocity field of gas or stars, measured on the receding side. 
For gas components, we include spaxels with H$\alpha$ S/N ratio greater than 3 in the fitting process, 
while for the stellar components, spaxels with the spectra median S/N ratio larger than 3 are used. 
The standard deviations of kinematic {\it PA}s of the gas ($\sigma_{\rm PA,gas}$) and stellar ($\sigma_{\rm PA,\star}$) components are calculated as 
\begin{equation} \label{equ:qlss}
  \sigma_{\rm PA} = \sqrt{\frac{1}{N} \sum_{i = 1}^{N}  (PA_i-\overline{PA} )^2}
\end{equation} 
where {\it PA}$_i$ is the {\it PA} at a certain radius and $\overline{PA}$ is the average value of {\it PA}$_i$ ($i$ = 1, 2, 3, ..., $N$, where $N$ is the number of radial bins). 
Objects with on-going merger features and irregular morphologies are not included in this study.

\noindent{\bf Kinematic model of the misaligned gas disks}\\
We model H$\alpha$ velocity fields as a combination of two rotating disks. 
The kinematic {\it PA} of the inner (outer) disk is defined as the median value inside (outside) the transition region.
We refer them as {\it PA}$_{\rm in}$ and {\it PA}$_{\rm out}$ as marked in Fig. 1c. 
We apply the relation $v_c(r)=Ar/(r^2+c_0^2)^{p/2}$ from the literature\cite{Bertola91} to fit the line-of-sight velocities $v_c$ (hereafter the velocity curve) as a function of radius $r$ along {\it PA}$_{\rm in}$ and {\it PA}$_{\rm out}$.
In the functional velocity curve model, $A$ is the maximum line-of-sight velocity, $c_0$ is the turnover radius of the velocity curve 
and $p$ represents the trend of the curve outside the turnover radius. 
We suppose $p=1$ for a flat velocity curve outside the turnover radius. 
Figure 6a shows the velocity curves along {\it PA}$_{\rm in}$ (orange dots) and {\it PA}$_{\rm out}$ (cyan dots). 
We assume the outer disk dominates at $r>r_2$, whereas the contribution from the inner disk can be neglected. 
The velocity curve of the outer disk is fitted based on velocities (cyan open dots) along {\it PA}$_{\rm out}$ outside a radius at which the velocity becomes flat.
At $r<r_1$, we fit data points (orange open dots) within the minimum and maximum velocities along {\it PA}$_{\rm in}$ to get the velocity curve of the inner disk. 
Considering that both the inner and outer disks contribute to the observed velocity curve at $r<r_1$, 
the intrinsic velocity curve for the inner disk along {\it PA}$_{\rm in}$ should be different from what we observe. 
Therefore, we set $A$ and $c_0$ of the inner disk, as well as $c_0$ of the outer disk, as free parameters, 
whereas $A$ of the outer disk is fixed at the maximum velocity along {\it PA}$_{\rm out}$ outside the turnover radius. 
The best-fit velocity curves of the inner and outer disks are shown as red and blue solid lines in Fig. 6a. 

The circular disk rotation fields\cite{Krajnovi06} are modelled based on the velocity curves as  
$v(r,\psi)=v_0+v_c(r)sin\ i\ cos\ \psi$, 
where $v_c(r)$ is the velocity curve, 
$v_0$ is the systemic velocity ($v_0=0\ {\rm km\ s^{-1}}$), 
and $\psi$ is the azimuthal angle measured from the major axis in the disk plane with a range of $\rm [-\pi/2, \pi/2]$. 
The inclination $i$ of the disk is defined as $cos\ i=b/a=q$, where $b$ and $a$ are the lengths of the minor and major axes. 
We set $q$ as a free parameter for both of the inner and outer disks. 
Figure 1e,h shows the velocity fields of the inner and outer gas disks. 

Based on the shape of the gas velocity field (Fig. 1c), 
it is clear that two galactic-scale gas components with different kinematics are required to model the velocity field.
In this work, we model the emission line at each spaxel by assuming two Gaussian components contributed by the inner and outer disks, respectively.
In Fig. 6b, the red and blue Gaussian profiles represent the contribution from the inner and outer disks at a certain spaxel, 
and the black profile is the combination of these two components, 
which models the observed H$\alpha$ emission line. 
We set the line centres of the two Gaussian components as the velocities of the inner and outer disks.  
The line widths are set to be the relevant value at the same spaxel of the velocity dispersion map. 
Our kinematic model is insensitive to the line width. 
The flux ratio of the two disks, $F_{\rm in}/F_{\rm tot}\equiv F_{\rm in}/(F_{\rm in}+F_{\rm out})$, is set as a free parameter, 
where $F_{\rm in}$ and $F_{\rm out}$ are the flux of the inner and outer disks at the relevant spaxel. 
We then fit one Gaussian (the green solid line in Fig. 6b) to the H$\alpha$ emission model to get the line centre. 
Figure 6c shows how the best-fit line centre changes with the flux ratio ($F_{\rm in}/F_{\rm tot}$). 
The best-fitted $F_{\rm in}/F_{\rm tot}$ (grey dashed line) is the value that gives the minimum value of $|v_{\rm obs}-v_{\rm mod}|$, 
where $v_{\rm obs}$ (magenta solid line) is the line centre of the observed $\rm H\alpha$ emission line and $v_{\rm mod}$ (black dots) is the line centre of the emission-line (EML) model given by the single Gaussian fit. 
Although $F_{\rm in}/F_{\rm tot}$ is set as a free parameter in our fitting process, the final result gives a smooth distribution of $F_{\rm in}/F_{\rm tot}$ over radius (Fig. 5b).

We also try to fit the observed H$\alpha$ emission lines with two Gaussian components; 
however, the reduced chi-squared $\chi^2$ of the double Gaussian fitting does not show obvious improvement 
(that is, a 15\% decrease of $\chi^2$) from the single Gaussian fitting. 
One possible reason that we cannot separate the two velocity components directly from the H$\alpha$ emission line 
is the limited spectral resolution of the MaNGA survey. 

We keep in mind that our model of gas kinematics is largely simplified since we totally ignore the interaction between the two accreted gas disks 
and the influence of the pre-existing gas component of the progenitor galaxy. 
We also assume the centres of the two gas disks overlap with each other, which is not necessarily true. 
However, this simplified kinematic model gives a reduced $\chi^2 < 2$ for these two objects, 
suggesting this model is a reasonable description of the observations. 

\end{methods}

\clearpage

\begin{figure*}[ht!]
    \centerline{ \includegraphics[width=0.9\textwidth]{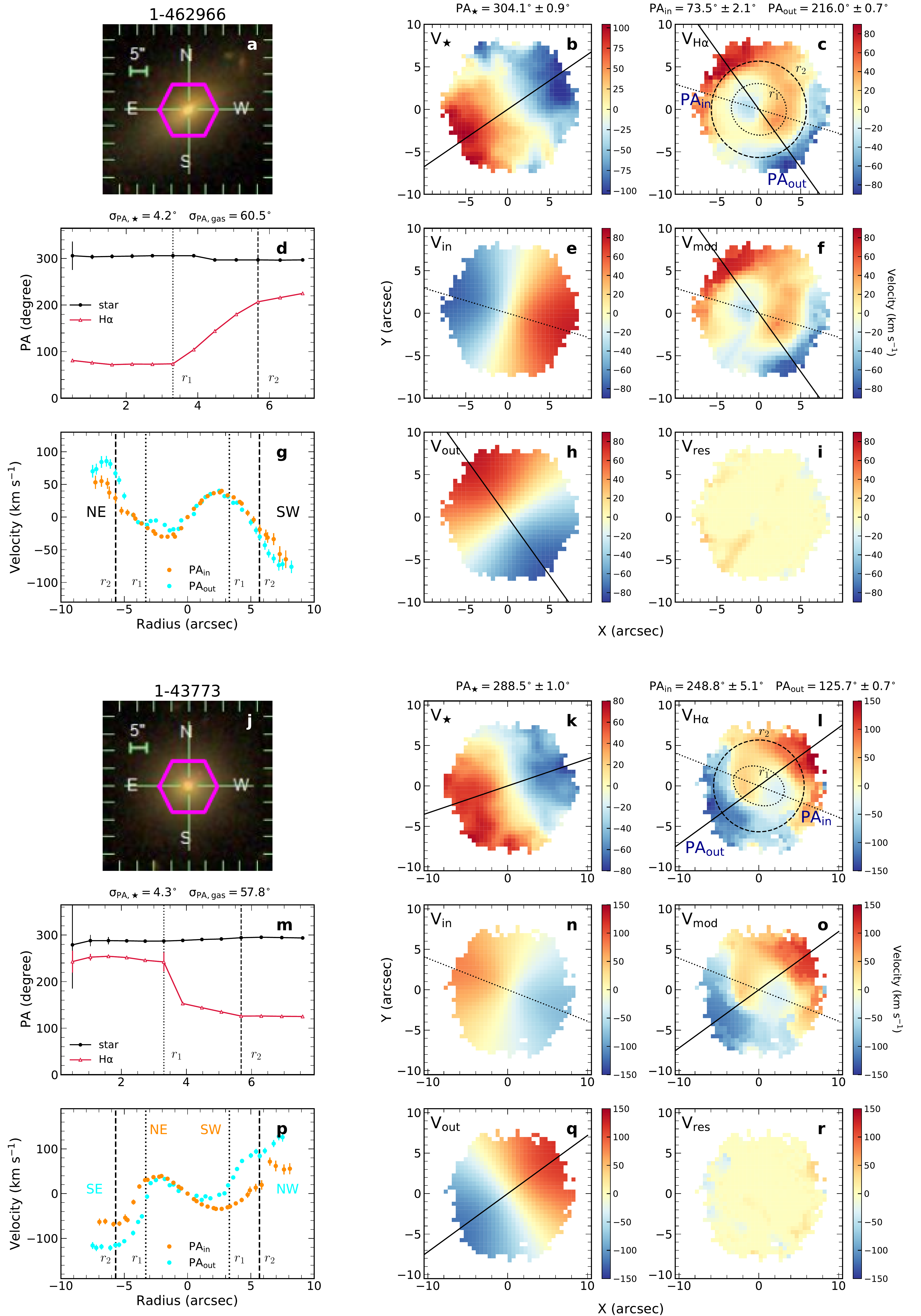}}
\end{figure*}
\clearpage
\begin{figure*}
    \caption{{\bf Kinematics of the two gas-gas misaligned galaxies.} 
    a-r, The first (a-i) and second (j-r) galaxies with the corresponding MaNGA-ID marked in a and j, respectively. 
    a,j, SDSS false-colour images. 
    b,c,k,l, The stellar ($V_\star$) (b,k) and gas ($V_{\rm H\alpha}$) (c,l) velocity fields. 
    The black solid line in b and k marks {\it PA}$_\star$. 
    The black dotted and solid lines in c and l mark {\it PA}$_{\rm gas}$ for the inner ({\it PA}$_{\rm in}$) and outer ({\it PA}$_{\rm out}$) gas disks. 
    The {\it PA} of the gas (red) and stellar (black) components are shown in d and m as a function of radius, 
    where the 2$\sigma$ uncertainties of the {\it PA} measurements have a median value of $\sim$4$^\circ$. 
    The black dotted and dashed lines labelled as $r_{\rm 1}$ and $r_{\rm 2}$ mark the transition region of {\it PA}$_{\rm gas}$, 
    which is also shown in c and l between the two ellipses. 
    e,h,n,q, The inner $V_{\rm in}$ (e,n) and outer $V_{\rm out}$ (h,q) gas disk models. 
    f,i,o,r, The best-fit gas velocity field $V_{\rm mod}$ (f,o) and the residual $V_{\rm res}$ (i,r); 
    the gas velocities along {\it PA}$_{\rm in}$ (orange dots) and {\it PA}$_{\rm out}$ (cyan dots) of the gas disks are shown in g and p, 
    where the error bars show 2$\sigma$ uncertainties, 
    and they have a median value of $\sim$5km s$^{-1}$.
  }
\end{figure*}

\begin{figure*}[ht!]
    \centerline{ \includegraphics[width=0.6\textwidth]{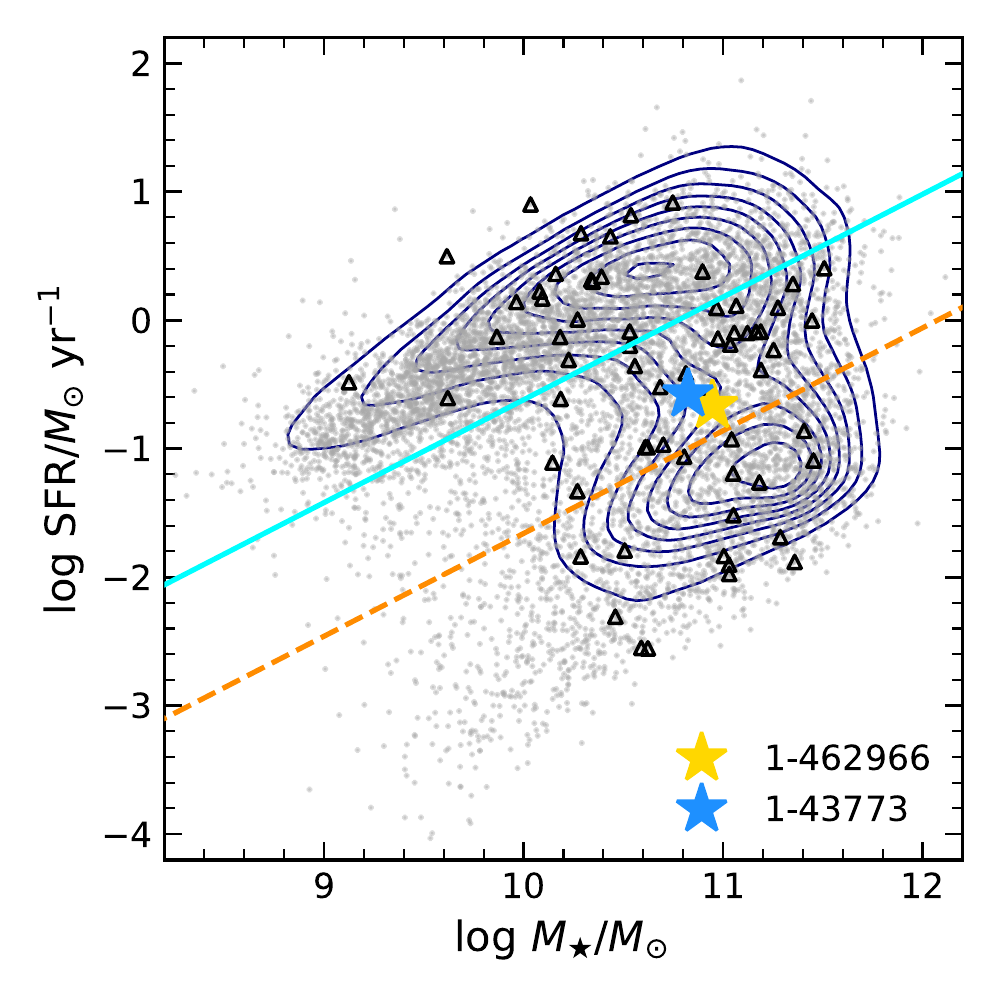}}
  \caption{{\bf SFRs versus stellar masses.} 
    The background contours show the GSWLC sample, while the grey dots are MaNGA galaxies.
    The open triangles are the 84 galaxies with disturbed gas kinematics, 
    and the stars show the two gas-gas misaligned galaxies studied in this work. 
    The two lines are used to separate galaxies into blue star-forming, GV and red quiescent. 
    The cyan solid line estimates the lower boundary of the star-forming galaxies (at the $\sim$1$\sigma$ level in scatter),
    the orange dashed line marks the upper limit of QS, 
    while galaxies between these two lines are classified as GV galaxies. 
}
\end{figure*}

\begin{figure*}[ht!]
    \centerline{ \includegraphics[width=0.85\textwidth]{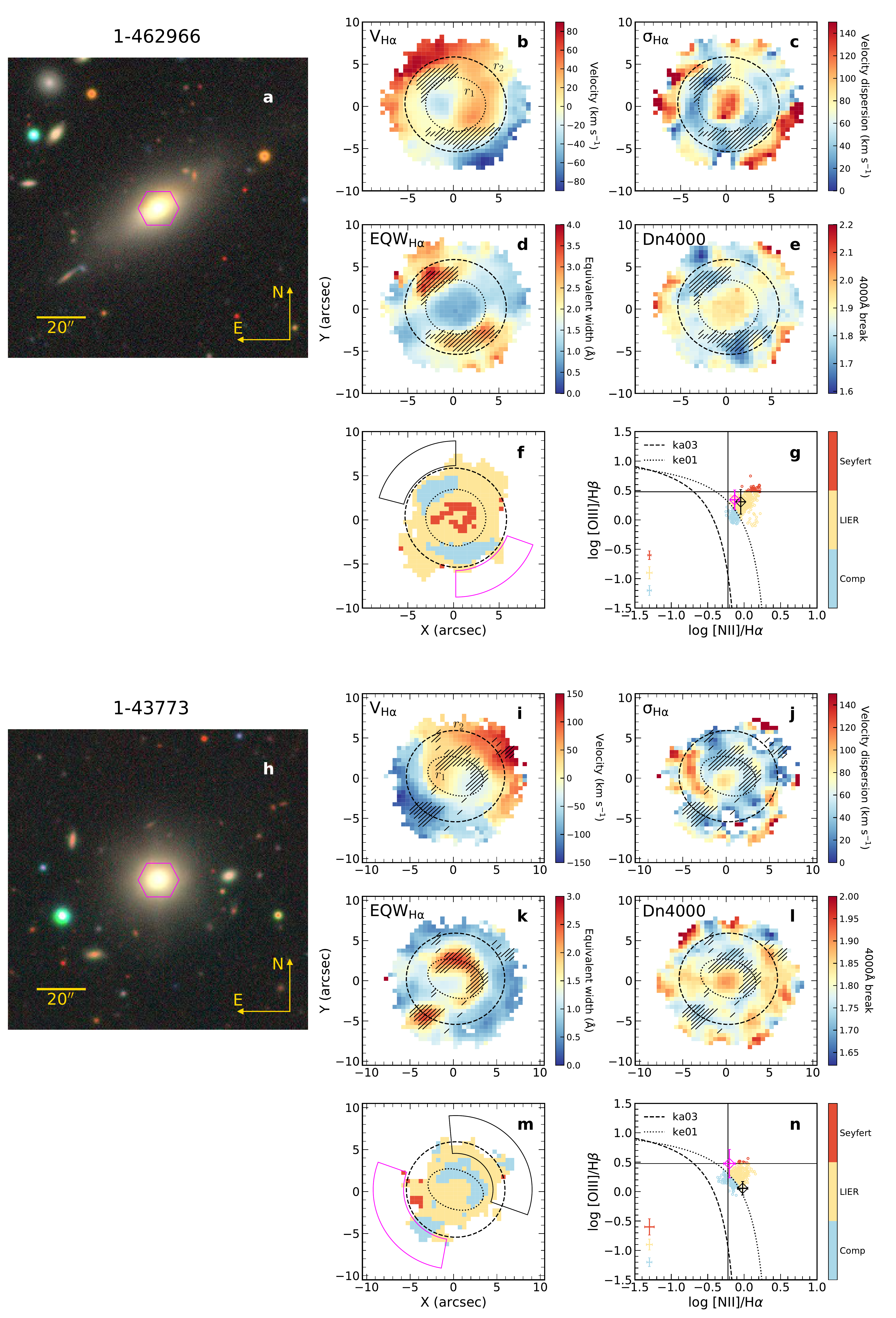}}
\end{figure*}
\clearpage
\begin{figure*}
    \caption{{\bf Enhanced star formation at the interface of the two gas disks.} 
    a-n, The first (a-g) and second (h-n) galaxies with the corresponding MaNGA-ID marked in a and g, respectively. 
    a,h, The $g$,$r$,$z$-band composite image from the DESI Legacy Survey. 
    b-e,i-l, The gas velocity field (b,i), gas velocity dispersion ($\sigma_{\rm H\alpha}$; c,j) H$\alpha$ EQW (d,k) and D$n$4000 (e,l) maps. 
    f,m, The spatially resolved BPT diagram. 
    g,n, The BPT diagnostic diagram\cite{Kauffmann03,Kewley01},
    where the dashed and dotted lines are the demarcations for star-forming (SF), composite (Comp., blue) and AGN (LIER, orange and Seyfert, red) regions. 
    The black and magenta diamonds show line ratios as well as their 2$\sigma$ uncertainties measured from stacked spectra of the outer gas disks. 
    The two stacked regions are marked in f and m. 
    The typical error bars shown in the bottom-left corner of g and n are the median of the 2$\sigma$ errors of Seyfert (red), LIER (orange) and composite (blue) spaxels, 
    where 39 Seyfert, 341 LIER and 91 composite spaxels in g, 
    and 11 Seyfert, 257 LIER and 109 composite spaxels in n.
}
\end{figure*}

\begin{figure*}[ht!]
    \centerline{ \includegraphics[width=0.8\textwidth]{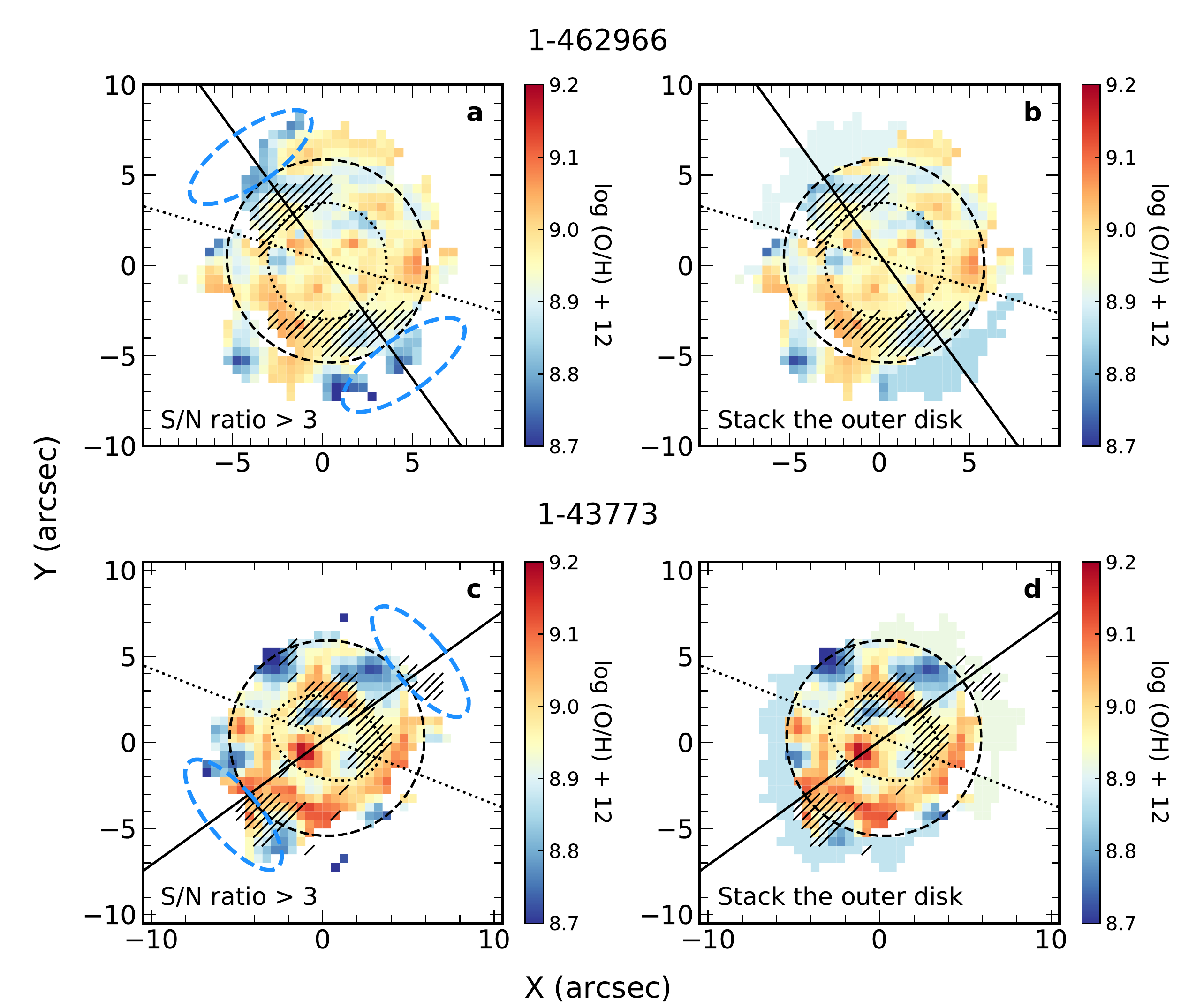}}
  \caption{{\bf Gas-phase metallicity of the two galaxies.} 
    a,c, The metallicity distributions for spaxels with \nii$\lambda$6585 and \sii$\lambda\lambda$6718,32 S/N ratio $>$ 3 for 1-462966 (a) and 1-43773 (c). 
    b,d, The same as in a,c, the only difference being that the gas-phase metallicity measured from stacked spectra for the outer disks is added. 
    The stacked regions for each galaxy are marked in the left panels as blue ellipses.
    }
\end{figure*}

\begin{figure*}[ht!]
    \centerline{ \includegraphics[width=1.\textwidth]{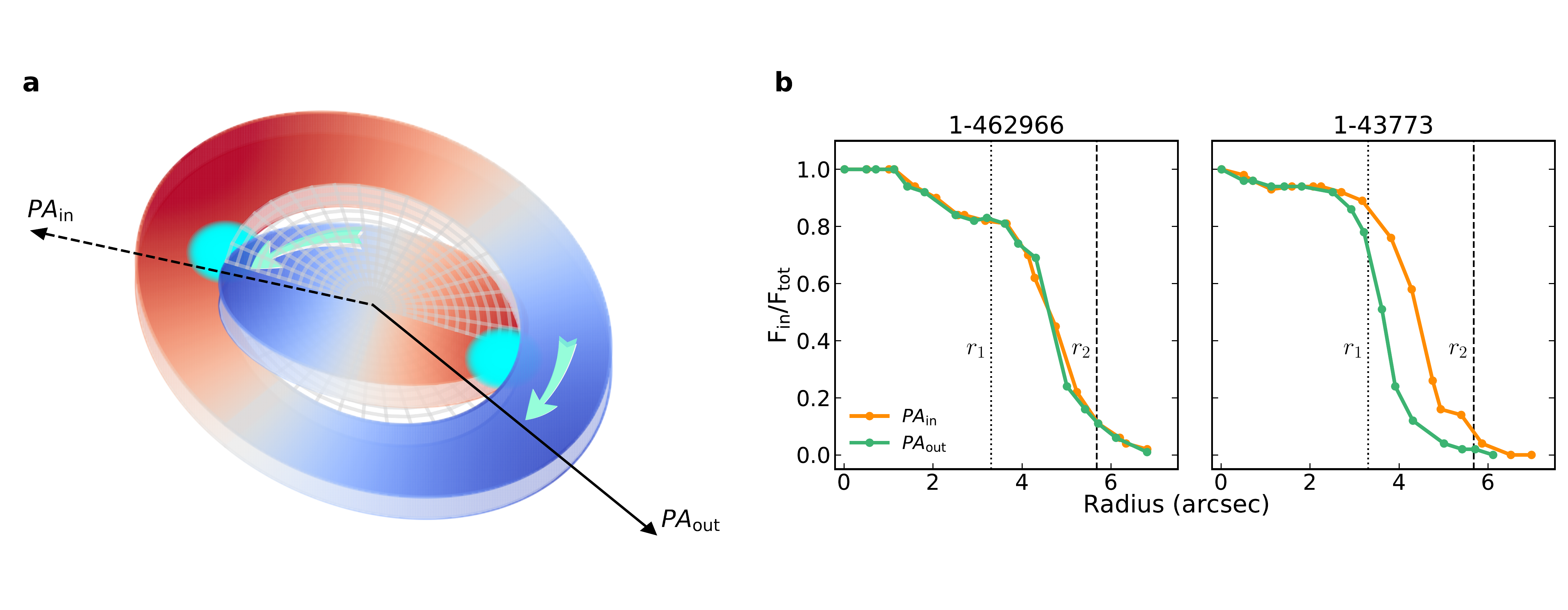}}
  \caption{{\bf Three dimensional configuration of two gas disks.} 
    a, A 3D cartoon of the inner and outer gas disks, with red and blue representing the gas component moving away from and towards us, respectively. 
    The green arrows show the rotation directions of the gas disks. 
    The dashed and solid lines are the kinematic major axes of the inner ({\it PA}$_{\rm in}$) and outer ({\it PA}$_{\rm out}$) disks. 
    The interface of the two disks is shown in cyan. b, The flux ratios ($F_{\rm in}/F_{\rm tot}$) as a function of radius along the kinematic major axes of the inner (orange) and outer (green) disks. 
    It is clear that the inner disk is dominant within $r_1$ while the outer disk is dominant outside $r_2$.
    }
\end{figure*}

\begin{figure*}[ht!]
    \centerline{ \includegraphics[width=1.\textwidth]{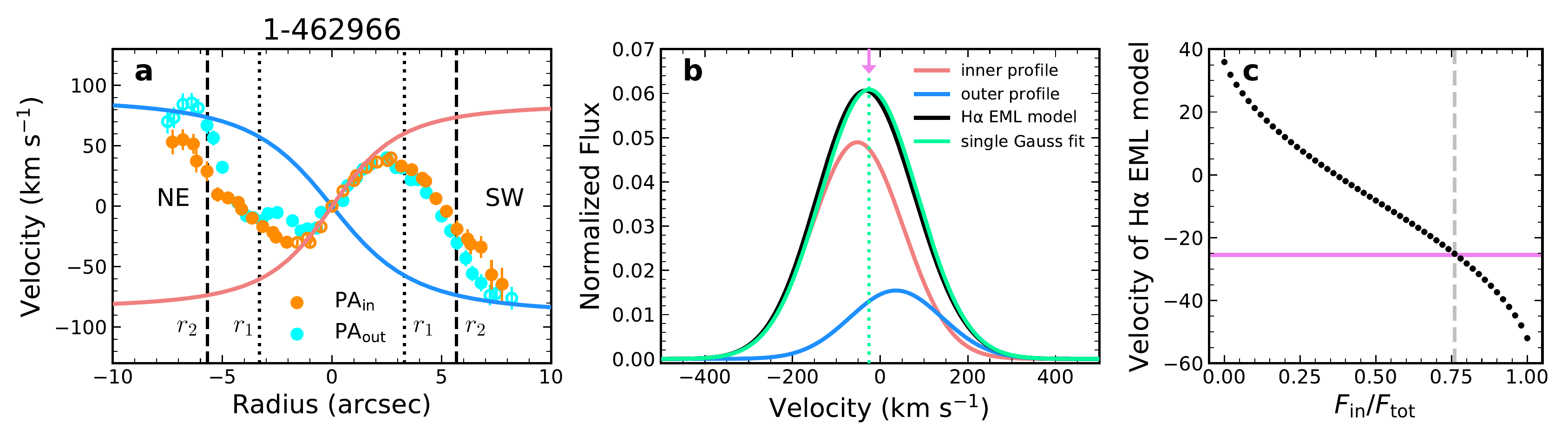}}
  \caption{{\bf Line-of-sight velocity curves and modelling the $\rm\bf\bm{H\alpha}$ emission line.} 
    a, Velocity curves and their 2$\sigma$ uncertainties along {\it PA}$_{\rm in}$ (orange dots) and {\it PA}$_{\rm out}$ (cyan dots) for the first galaxy, 
    where the transition region between $r_1$ and $r_2$ is marked with vertical dotted and dashed lines. 
    The best-fit velocity curves along {\it PA} for the inner and outer disks are shown as red and blue solid lines, respectively. 
    b, The red and blue Gaussian profiles representing the contribution of the inner and outer disks at a certain spaxel. 
    The combination of the red and blue components shown as the black line is the H$\alpha$ EML model. 
    A single Gaussian profile is used to fit the H$\alpha$ EML model, the best fit is shown in green with the line centre marked as green dotted line and the magenta arrow indicates the observed velocity at this spaxel. 
    c, The best-fit line centre of the EML model as a function of $F_{\rm in}/F_{\rm tot}$. 
    The best-fit $F_{\rm in}/F_{\rm tot}$ (grey dashed line) is the value that gives the minimum of $|v_{\rm obs}-v_{\rm mod}|$, 
    where $v_{\rm obs}$ (magenta solid line) is the line centre of the observed H$\alpha$ emission line and $v_{\rm mod}$ (black dots) is the best-fit line centre of the EML model.
    }
\end{figure*}

\clearpage

\noindent{\bf Data availability}\\
The MPL-11 data (including all the MPL-10 data) that support the findings of this paper are available through SDSS Data Release 17 which can be downloaded from https://www.sdss.org/dr17/manga/. 
The multi-waveband images from DESI Legacy Surveys can be downloaded from https://www.legacysurvey.org/. 

\noindent{\large\bf References}

\bibliographystyle{naturemag,aastex}

\begin{addendum}

\item[Acknowledgements]
We thank D. Xu, S. Feng, T. Wang and M. Xiao for helpful discussion and comments. 
Y.M.C. acknowledges support from the National Key Research and Development Program of China (grant No. 2017YFA0402700), 
the National Natural Science Foundation of China (grant Nos. 11922302, 11733002 and 12121003), 
and the China Manned Space Project (grant No. CMS-CSST-2021-A05). 
A.M. acknowledges support from the Special Astrophysical Observatory of the Russian Academy of Sciences government contract approved by the Ministry of Science and Higher Education of the Russian Federation. 
D.B. acknowledges a partial support from the Russian Science Foundation (grant No. 22-12-00080). 
J.G.F-T. acknowledges support from Proyecto Fondecyt Iniciaci\'on (grant No. 11220340), ANID Concurso de Fomento a la Vinculaci\'on Internacional para Instituciones de Investigaci\'on Regionales Proyecto (grant No. FOVI210020), and the Joint Committee ESO-Government of Chile 2021 (grant No. ORP 023/2021). 
R.R. acknowledges support from 
Conselho Nacional de Desenvolvimento Cient\'{i}fico e Tecnol\'ogico (CNPq, Proj. 311223/2020-6, 304927/2017-1 and 400352/2016-8), 
Funda\c{c}\~ao de amparo \`{a} pesquisa do Rio Grande do Sul (FAPERGS, Proj. 16/2551-0000251-7 and 19/1750-2), 
and Coordena\c{c}\~ao de Aperfei\c{c}oamento de Pessoal de N\'{i}vel Superior (CAPES, Proj. 0001).
Funding for the Sloan Digital Sky 
Survey IV has been provided by the 
Alfred P. Sloan Foundation, the U.S. 
Department of Energy Office of 
Science, and the Participating 
Institutions. SDSS-IV acknowledges support and 
resources from the Center for High 
Performance Computing  at the 
University of Utah. The SDSS 
website is www.sdss.org.
SDSS-IV is managed by the 
Astrophysical Research Consortium 
for the Participating Institutions 
of the SDSS Collaboration including 
the Brazilian Participation Group, 
the Carnegie Institution for Science, Carnegie Mellon University, Center for 
Astrophysics | Harvard \& 
Smithsonian, the Chilean Participation 
Group, the French Participation Group, 
Instituto de Astrof\'isica de 
Canarias, The Johns Hopkins 
University, Kavli Institute for the 
Physics and Mathematics of the 
Universe (IPMU) / University of 
Tokyo, the Korean Participation Group, 
Lawrence Berkeley National Laboratory, 
Leibniz Institut f\"ur Astrophysik 
Potsdam (AIP),  Max-Planck-Institut 
f\"ur Astronomie (MPIA Heidelberg), 
Max-Planck-Institut f\"ur 
Astrophysik (MPA Garching), 
Max-Planck-Institut f\"ur 
Extraterrestrische Physik (MPE), 
National Astronomical Observatories of 
China, New Mexico State University, 
New York University, University of 
Notre Dame, Observat\'ario 
Nacional / MCTI, The Ohio State 
University, Pennsylvania State 
University, Shanghai 
Astronomical Observatory, United 
Kingdom Participation Group, 
Universidad Nacional Aut\'onoma 
de M\'exico, University of Arizona, 
University of Colorado Boulder, 
University of Oxford, University of 
Portsmouth, University of Utah, 
University of Virginia, University 
of Washington, University of 
Wisconsin, Vanderbilt University, 
and Yale University.

\item[Author contributions]
X.C. made the plots and led the writing of the very preliminary version of the draft. Y.M.C. discovered these two galaxies and conceived the project. Y.M.C. and Y.S. suggested the physical pictures discussed in this paper and edited the manuscript.
M.B. helped in fitting the disk models. 
A.M., D.B., S.-L.L., J.G.F.-T., R.A.R., R.R. and R.R.L. were involved in the comments in the manuscript and the interpretation of the results.

\item[Competing Interests]
The authors declare no competing interests.

\item[Additional information]
Correspondence and requests for materials should be addressed to Y.M.C. 
(Email: chenym@nju.edu.cn)

\end{addendum}

\end{document}